\documentclass[aps,prl,twocolumn,groupedaddress]{revtex4}

\begin{document}

\title{Approach to Physical Reality: a note on Poincar\'{e} Group \\and the philosophy of Nagarjuna.}

\author{David Vernette}
\author{Punam Tandan}
\author{Michele Caponigro}

\affiliation{}

\emph{\date{\today}}
\\
\\
\begin{abstract}
We argue about a possible scenario of physical reality based on
the parallelism between Poincar\'{e} group and the sunyata
philosophy of Nagarjuna. The notion of "relational" is the common
denominator of two views. We have approached the relational
concept in third-person perspective (ontic level). It is possible
to deduce different physical consequence and interpretation
through first-person perspective approach. This relational
interpretation leave open the questions: i)we must abandon the
idea for a physical system the possibility to extract the
completeness information? ii)we must abandon the idea to infer a
possible structure of physical reality?

\end{abstract}

\maketitle

\section{Poincar\'{e} Group}
There are two universal features of modern day physics regarding
physical systems: all physical phenomena take place in
1)space-time and all phenomena are (in principle) subject to
2)quantum mechanics. Are these aspects just two facets of the same
underlying physical reality? The research is concentrate on this
fundamental point. The notion of space-time is linked to the
geometry, so an interesting question is what geometry is
appropriate for quantum physics\footnote{note -1- It was
suggested, for instance, that the universal symmetry group
elements which act on all Hilbert spaces may be appropriate for
constructing a physical geometry for quantum theory.}. Can
geometry give us any knowledge about the nature of physical space
where the physical laws take place? Can geometry give us the
possible scenario of the physical reality? A fundamental aspect of
a geometry is the group of transformations defined over it. Group
theory is the necessary instruments for expressing the laws of
physics (the concept of symmetry is derived from group
theory.)\footnote{note -2- Some authors retain the symmetry is the
ontic element, and the physical laws like the space-time are
secondary}.Physics and the geometry in which it take place are not
independent. We retain there is a close relationship between
space-time structure and physical theory. Space-time imposes
universally valid constraints on physical theories and the
universality of these laws starts to become less mysterious (i.e.
various paradox). The invariance under the group of
transformations is a fundamental criterion to classify
mathematical structures. Poincar\'{e} introduced notion of
invariance under continue transformations. The Poincar\'{e} Group
is the group of translation, rotation, and boost operators in
4-dimensional space-time. Now, some natural questions are:
\textbf{does space exist independently of phenomena?} Itself has
an intrinsic significance? A system defined in this space through
physical law could exist by itself? We call "absolute" reality the
reality of a system that do not depend by its interaction with
other system. The problem is that we have not a single system. In
this brief note, we abandon the idea of absolute reality and we
argue in favor of a relational reality, because relational reality
is founded on the premise that an \textbf{object is real only} in
relation to another object that it is interacting with. In the
relational interpretation\cite{ro}, the basic elements of
objective reality are the measurement events themselves. This
interpretation goes beyond the Copenhagen interpretation by
replacing the \textbf{absolute reality with relational reality}.
In the relational interpretation the wave function is merely a
useful mathematical abstraction. Some authors proposes that the
laws of nature are really the result of probabilities constrained
by fundamental symmetries. \textbf{Relational reality is
associated with the fundamental concept of interactions.} These
later analysis of the "relational" notion bring us to approach the
same problem utilizing the sunyata philosophy of Nagaujuna.

\section{Concept of reality in the philosophy of Nagarjuna}

The Middle Way of Madhyamika refers to the teachings of Nagarjuna,
very interesting are the implications between quantum physics and
Madhyamika. The basic concept of reality in the philosophy of
Nagarjuna is that the fundamental reality has no firm core but
consists of systems of \textbf{interacting objects}. According to
the middle way perspective, based on the notion of emptiness,
phenomena exist in a relative way, that is, they are empty of any
kind of inherent and independent existence. Phenomena are regarded
as dependent events existing \textbf{relationally rather than
permanent things}, which have their own entity. Nagarjuna middle
way perspective emerges as a relational approach, based on the
insight of emptiness. Sunyata (emptiness) is the foundation of all
things, and it is the basic principle of all phenomena. The
emptiness implies the negation of unchanged, fixed substance and
thereby the possibility for relational existence and change. This
suggests that both the ontological constitution of things and our
epistemological schemes are just as relational as everything else.
We are fundamentally relational internally and externally. In
other words, Nagarjuna do not fix any ontological nature of the
things:
\begin{itemize}
\item 1)they do not arise.
\item 2)they do not exist.
\item 3)they are not to be found.
\item 4)they are not.
\item 5)and they are unreal
\end{itemize}
In short, an invitation do not decide on either existence or
non-existence(nondualism). According the theory of sunyata,
phenomena exist in a relative state only, a kind of
'\textbf{ontological relativity}'. Phenomena are regarded as
dependent(only in relation to something else) events rather than
things which have their own inherent nature; thus the extreme of
permanence is avoided.

\section{Conclusion}
We have seen the link between \textbf{relational and interaction}
within a space-time governed by own geometry. Nagarjuna's
philosophy use the same basic concept of "relational" in the
interpretation of reality. We note that our parallelism between
the scenario of physical reality and the relational interpretation
of the same reality is based on third-person perspective approach
(i.e. the ontic level, relational view include the
observer-device). Different considerations could be done thought
first-person perspective approach, in this case we retain the
impossibility to establish any parallelism. Finally, we note that
probably the relational approach stimulate the interest to
fundamental problems in physics like: the unification of laws and
the discrete/continuum view\cite{da}.

\section{{\tiny  }}
{\footnotesize------------------\\ $\diamond$\emph{David Vernette, Punam Tandan, Michele Caponigro}\\
\\
Quantum Philosophy Theories
www.qpt.org.uk\\
\\
$\diamond$ \emph{qpt@qpt.org.uk} }
\\
\\

\end{document}